\documentclass[%
 aip,
 amsmath,amssymb,
 reprint,%
]{revtex4-1}

\usepackage{graphicx}
\usepackage{dcolumn}
\usepackage{amsmath,bm}

\usepackage[utf8]{inputenc}
\usepackage[T1]{fontenc}
\usepackage{mathptmx}
\usepackage{etoolbox}
\usepackage{comment}
\usepackage{xcolor}

\newcolumntype{C}[1]{>{\centering\arraybackslash}m{#1}}

\usepackage[normalem]{ulem}

\makeatletter
\def\@email#1#2{%
 \endgroup
 \patchcmd{\titleblock@produce}
  {\frontmatter@RRAPformat}
  {\frontmatter@RRAPformat{\produce@RRAP{*#1\href{mailto:#2}{#2}}}\frontmatter@RRAPformat}
  {}{}
}%

\makeatother
\begin{document}

\preprint{AIP/123-QED}

\title{Electric field control for experiments with atoms in Rydberg states}

\author{Aishik Panja}
\affiliation{ 
Department of Physics and Astronomy, Purdue University, West Lafayette, IN 47907, USA
}

\author{Yupeng Wang}%
\affiliation{ 
Department of Physics and Astronomy, Purdue University, West Lafayette, IN 47907, USA
}%

\author{Xinghan Wang}
\affiliation{ 
Department of Physics and Astronomy, Purdue University, West Lafayette, IN 47907, USA
}%
\author{Junjie Wang}
\affiliation{ 
Department of Physics and Astronomy, Purdue University, West Lafayette, IN 47907, USA
}
\affiliation{ 
School of the Gifted Young, University of Science and Technology of China, Hefei 230026, China
}
\author{Sarthak Subhankar}
\affiliation{ 
Joint Quantum Institute, National Institute of Standards and Technology and the University of Maryland, College Park,
 MD 20742, USA
}%

\author{Qi-Yu Liang}
\affiliation{ 
Department of Physics and Astronomy, Purdue University, West Lafayette, IN 47907, USA
}%
\affiliation{Purdue Quantum Science and Engineering Institute, Purdue University, West Lafayette, IN 47907, USA}

\email{qyliang@purdue.edu}
\date{\today}

\begin{abstract}
Atoms excited to Rydberg states have recently emerged as a valuable resource in neutral atom platforms for quantum computation, quantum simulation, and quantum information processing. Atoms in Rydberg states have large polarizabilities, making them highly sensitive to electric fields. Therefore, stray electric fields can decohere these atoms, in addition to compromising the fidelity of engineered interactions between them. It is therefore essential to cancel these stray electric fields. Here we present a novel, simple, and highly-compact electrode assembly, implemented in a glass cell-based vacuum chamber design, for stray electric field cancellation. The electrode assembly allows for full 3D control of the electric field in the vicinity of the atoms while blocking almost no optical access. We experimentally demonstrate the cancellation of stray electric fields to better than 10 mV/cm using this electrode assembly.

\end{abstract}

\maketitle

\section{\label{sec:Introduction}Introduction}


Rydberg atoms play a crucial role in advancing quantum information processing~\cite{Morgado2021,browaeys2020many,bluvstein2024logical}, quantum optics~\cite{firstenberg2016nonlinear,shao2024rydberg} and quantum sensing~\cite{meyer2020assessment,fancher2021rydberg}. Notably, their $n^4$ and $n^{11}$ scaling ($n$ being the principal quantum number) for the strength of dipole-dipole interaction and Van-der-Waals interaction provide neutral atoms with strong mutual interactions that are otherwise challenging to achieve. Additionally, their $n^7$ scaling for polarizability can be leveraged to build the next generation of electric field sensors.
However, this extreme sensitivity also makes Rydberg atoms vulnerable to stray electric fields, which can be detrimental in Rydberg-interaction mediated quantum computation and quantum simulation experiments~\cite{RevModPhys.82.2313, evered2023high}. While electrodes can be built to cancel these stray electric fields~\cite{peyronel2013quantum,de2018quantum,ornelas2020experiments,mirgorodskiy2017storage,lorenz2021rydberg}, such designs must not restrict optical access, especially the high numerical aperture (NA) optical access typically required by these experiments.

Glass cell-based vacuum chamber designs typically provide the highest amount of optical access compared to stainless steel vacuum chamber designs. However, the inner diameter of the ConFlat flange that connects the glass cell to an auxiliary chamber is small. Therefore, engineering highly compact electrode structures that can cancel out stray electric fields without compromising the large optical access of glass cells is a challenging task. 

There are a few strategies for controlling the electric field in the vicinity of the atoms: in-vacuo Faraday cage~\cite{anand2024dual}, manufacturing electrodes with the glass cells~\cite{saffmanelelectricfieldcontrolpatent}, ultrahigh vacuum-compatible printed circuit boards~\cite{wilson2022trapping,wilson2022new}, and placing electrodes outside the vacuum system~\cite{archimi2022measurements}. While effective in canceling stray electric fields, these designs have a few shortcomings: large electrode structures can be hard to assemble inside vacuum and can compromise the ability to adapt to changes in optical layouts; the electrode structure can act as a cavity and can induce blackbody-radiation-induced stimulated emission of atoms in Rydberg states effectively reducing the Rydberg state lifetimes; Faraday cages, apart from restricting optical access, also decrease the conductance to the vacuum pumps and therefore reduce the vacuum limited-lifetime of the atoms; electrodes placed outside glass cells typically requires higher voltages and often results in larger residual stray fields.

Here we report on a novel, simple, and highly compact electrode assembly design that allows us to fully control the electric field in the vicinity of the atoms without restricting high NA optical access. We achieve instantaneous stray electric
field cancellation to better than 10 mV/cm, with drifts of no more than 20 mV/cm over a few hours and 50 mV/cm day-to-day. This level of electric control is essential for atoms excited to high $n$ or high angular momentum $l$ Rydberg states. Our design can be implemented in almost any glass cell with little to no modifications.




\section{\label{sec:Electrode design}Electrode design}
\subsection{\label{sec:sec:Mechanical design}Mechanical design}

The experiments were conducted using a UV fused silica octagonal glass cell from Precision Glassblowing, which has a 2-3/4" ConFlat flange and no anti-reflection coated or nanotextured surfaces. The electrodes are thin tungsten rods (diameter: 0.381 mm) that are supported by Macor and Alumina structures. Furthermore, the materials surrounding the atomic cloud strongly affect the stray electric fields generated in the vicinity of the atoms~\cite{levine2018high,wilson2022trapping}. The nearest dielectric surface, the glass cell, is 11 mm away. The center of the three dispensers (RB/NF/3.4/12
FT10+10, SAES Getters) is $\sim40$~mm away. 
Alumina rods (Al2O3-TU-C-3000, Kimball Physics) provide mechanical supports and are positioned $\sim50$~mm away from the atoms.

\begin{figure}[ht]
\centering
\includegraphics[width = \columnwidth]{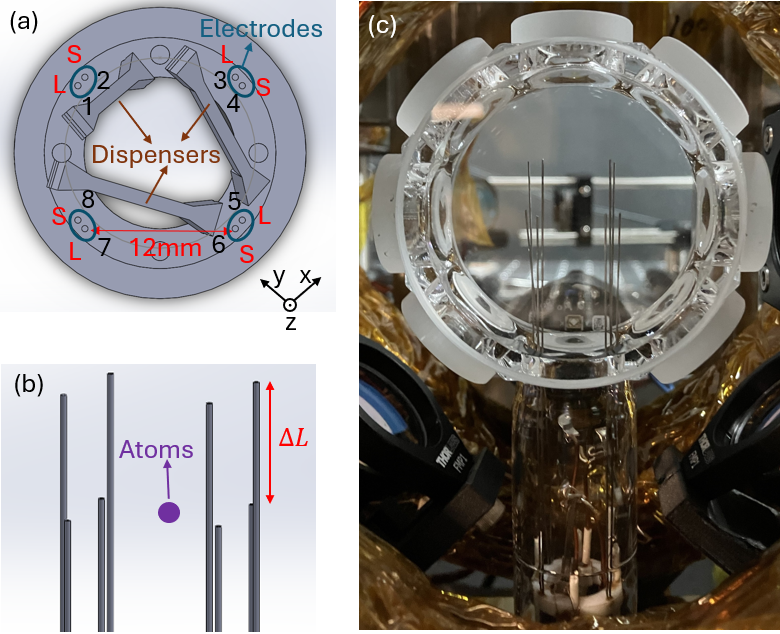}
\caption{Electrode configuration: (a) top view and (b) angled side view from the CAD file. `L' and `S' indicate the long and short electrodes, respectively. The long electrodes are $\Delta L=$1~cm longer than the short ones. The atomic cloud should be prepared in the geometric center of the electrodes, aligned with the top of the short electrodes. (c) Photo of the octagonal glass cell with the electrodes inside. 
}
\label{fig:electrode_geometry}
\end{figure}

To realize full electric field control, including the vector field and its spatial derivatives, 8 electrodes are required. The electrodes are arranged in a square pattern (numbered 1 to 8 in Fig.~\ref{fig:electrode_geometry}a). Each corner of the square features one long and one short rod, with a length difference of $\Delta L =10$~mm. The spacing between the corners of the square pattern is limited by the smallest inner diameter along the neck of the glass cell (19~mm). The homogeneity of the electric field in the vicinity of the atomic cloud increases with the spacing. The gap between each long rod and its neighbouring short rod is kept to a minimum while still avoiding electrical shorts.


\subsection{\label{sec:Simulation of performances}Simulation of performances}

During the ultrahigh vacuum chamber bake-out, electrodes 6 and 7 (see Fig.~\ref{fig:electrode_geometry}(a) for the electrode numbering scheme) got shorted. Despite losing one degree of freedom from the short, we can still control all necessary parameters if not all degrees of freedom are needed simultaneously. To cancel the ambient electric field, we apply voltages as specified in Table \ref{tab:how_to_apply_E_field}. 
We use COMSOL AC/DC module to estimate the strength of applied electric field and its inhomogeneity. We include the glass cell in our simulations. The glass cell is made out of Corning 7980 glass and has a relative permittivity $\epsilon_r=3.8$. The axes in the inset of Fig.~\ref{fig:electrode_geometry}(a) result from simulating the configuration specified in Table \ref{tab:how_to_apply_E_field}.  
According to the simulation, $V_x$ $(V_y)$ on electrode 8 (5) and $-V_x$ $(-V_y)$ on electrode 4 (1) generates electric field almost entirely along the $x$ $(y)$ axis. Similarly, $-V_z$ on the long electrodes $1\And5$ and $V_z$ on the short electrodes $4\And8$ generates an electric field along $z$. These voltages are linearly combined (Table \ref{tab:how_to_apply_E_field}) to generate an arbitrary electric field. 
\begin{table}[]
    \centering
\begin{tabular}{|C{1.2cm}|C{1.2cm}|C{1.2cm}|C{1.2cm}|}
     \hline
     $V_1$ & $V_4$ & $V_5$ & $V_8$ \\
     \hline
     $-V_y-V_z$ & $-V_x+V_z$ & $V_y-V_z$ & $V_x+V_z$\\
     \hline
\end{tabular}
    
    \caption{Electrode configuration to cancel stray electric fields. Voltages are applied on electrodes 1, 4, 5 and 8, denoted as $V_1$, $V_4$, $V_5$ and $V_8$. The other four electrodes not in the table are grounded. The electrode numbering is shown in Fig.~\ref{fig:electrode_geometry}(a).}
    \label{tab:how_to_apply_E_field}
\end{table}



\section{Characterization of electric field control}
\label{sec:Characterization of electric field control}

\subsection{Rydberg spectroscopy}

\begin{figure*}
\centering
\includegraphics[width=0.8\textwidth]{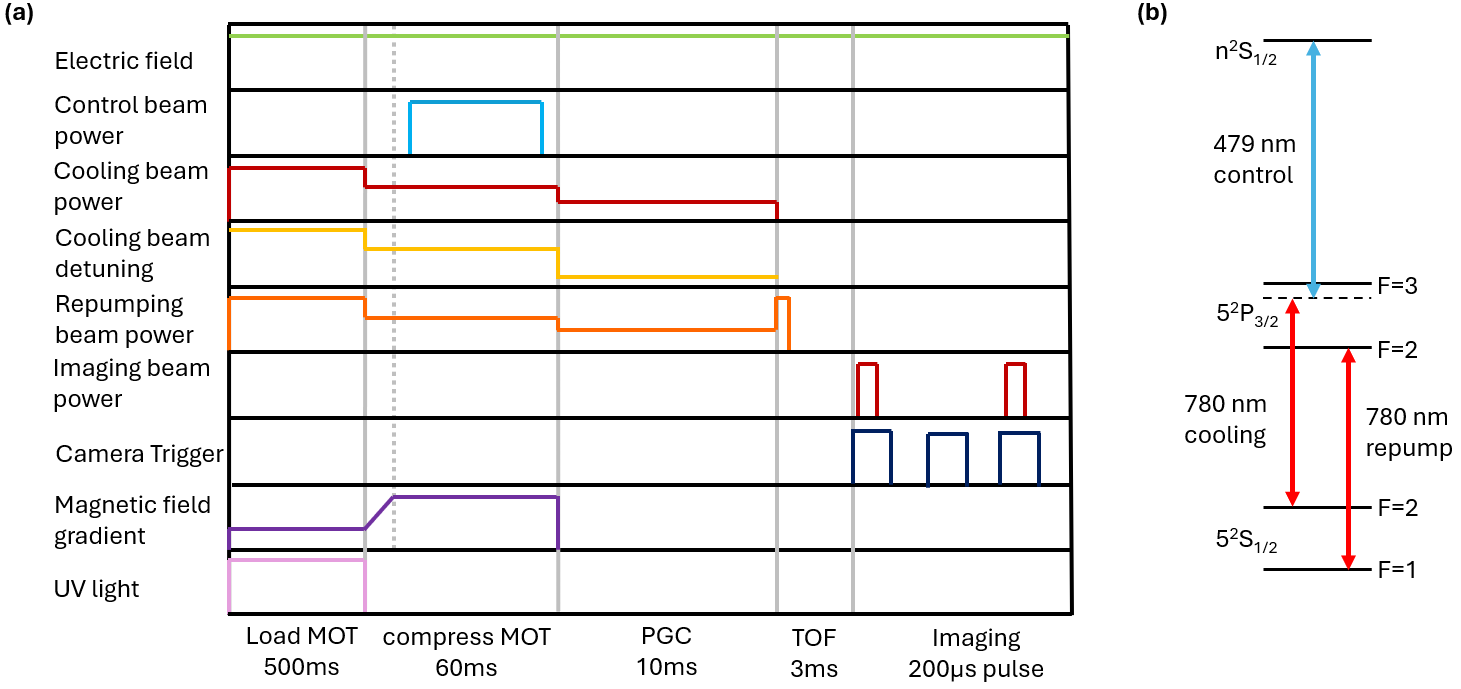}
\caption{Schematic of the (a) experimental sequence (not to scale) and (b) the level structure.}
\label{fig:experiment_sequence}
\end{figure*}

First, we describe a typical experimental sequence. We employ light-induced atomic desorption (LIAD)~\cite{torralbo2015light,klempt2006ultraviolet} to desorb $^{87}$Rb from the inner surface of the glass cell to load a 3D magneto-optical trap (MOT). Two $~1$~W ultraviolet (UV) LEDs (365 nm, Thorlabs M365L3) illuminate the glass cell during the 0.5~s MOT loading stage.  All dispensers are grounded during the experiments. To prevent Rb atom depletion from the glass cell walls during consecutive days of experimentation, we maintain a low current flow through one of the dispensers overnight. We collect roughly $10^5$ atoms in the MOT using three retro-reflected overlapping cooling and repumping beams with $1/e^2$ radius of $8.8$~mm. At this stage, the total power of the three cooling and repumping beams is 100~mW and 9~mW, respectively. We compress the MOT by ramping up the magnetic field gradient from 27.5 to 54~G/cm along the anti-Helmholtz gradient coil axis in 10~ms and holding the atoms in the high field gradient for 50~ms. Then the atoms are cooled to 35~$\mu$K using polarization gradient cooling (PGC). We take absorption imaging of the atomic cloud after 3~ms time-of-flight (TOF). Three images are taken consecutively, denoted as atoms (a), dark (d) and probe (p) and separated in time by 13.7~ms. Optical depth (OD) is calculated as OD$\equiv -\text{ln}\cfrac{I_a(x,y) - I_d(x,y)}{I_p(x,y) - I_d(x,y)}$. We sum over all the column OD within a region of interest to get the integrated OD.

\begin{figure*}
\centering
\includegraphics[width=0.68\textwidth]{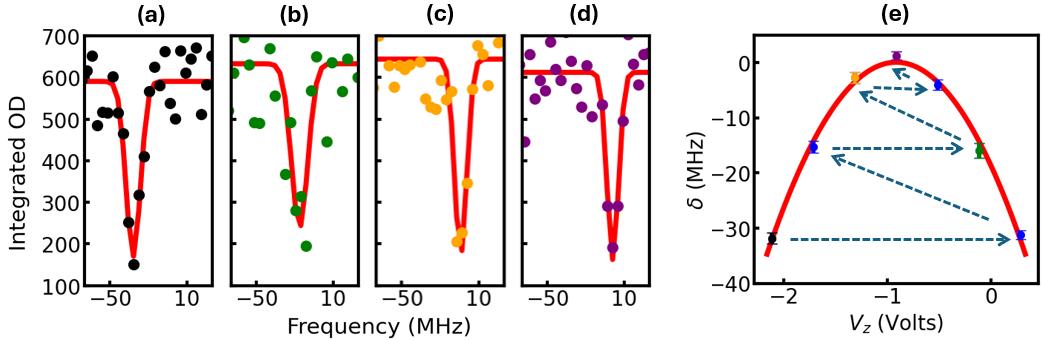}
\caption{Rydberg spectroscopy. The red lines in (a-d) are Gaussian fits to the spectroscopy data, obtained by scanning the control laser frequency. The center of each Gaussian is marked as $\delta$ 
in (e). The dashed blue arrows in (e) labels the order of data points taken. The black, green, yellow and purple dots in all the plots correspond to 
332(3), 209(3), 263(3) and 237(3)
mV/cm, respectively. The red line represents a parabolic fit based on Eq.~\ref{eq2}.}
\label{fig:Rydberg_spectroscopy}
\end{figure*}

We modify the sequence above to perform Rydberg spectroscopy using a two-photon process during the compressed MOT (cMOT) stage (see Fig.~\ref{fig:experiment_sequence}(a) for a graphic presentation of the experimental sequence). The detuning of the cooling beam is -15 MHz during the `Load MOT' stage, which is changed to -75 MHz in the cMOT stage for Rydberg spectroscopy.  Additionally, we apply a control beam at 479 nm (Fig.~\ref{fig:experiment_sequence}(b)) with $1/e^2$ radius of 20~$\mu$m for 40~ms.
  At this stage, the power of the cooling and control beams is 80~mW and 15~mW, respectively. These beams induce Raman excitation to Rydberg states, depleting the atoms from the cMOT. For spectroscopy of the Rydberg states, we tune the frequency of the control beam. Additional details can be found in Appendix A. 
  


\begin{figure*}
\centering
\includegraphics[width=0.7\textwidth]{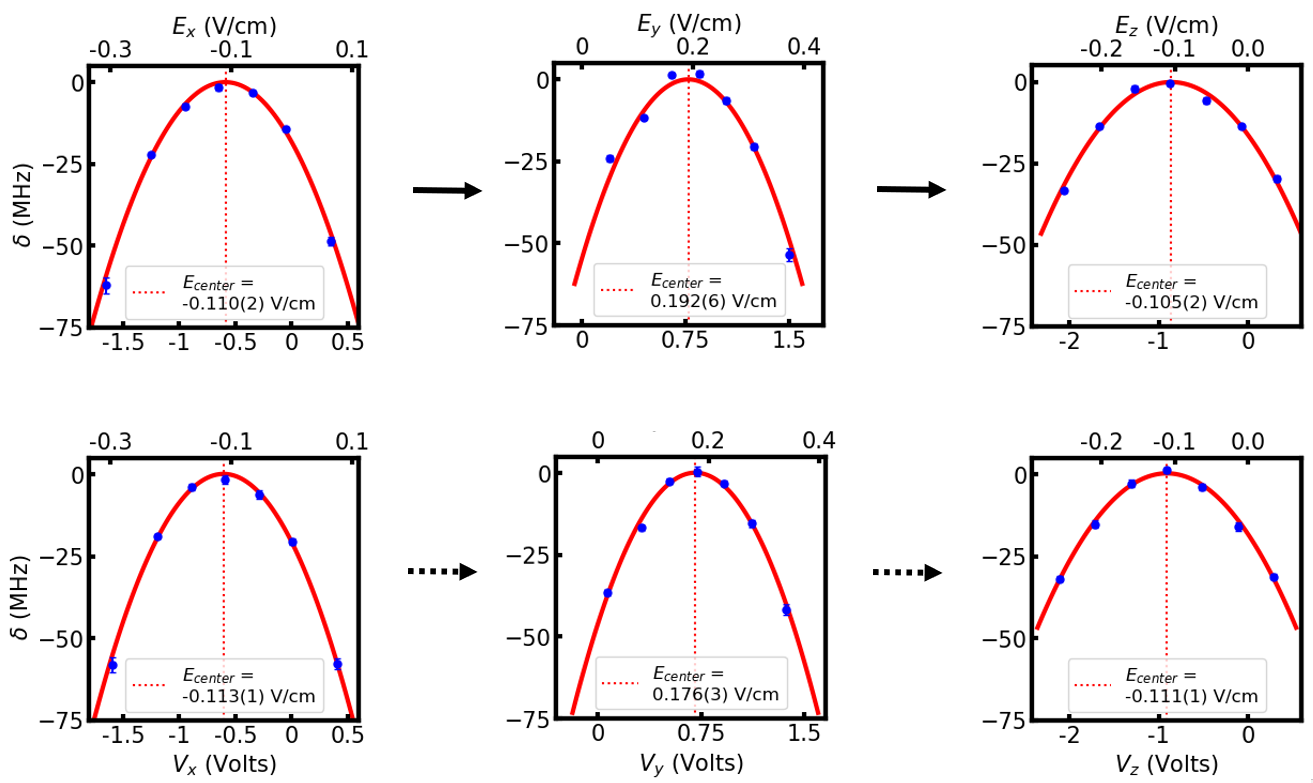}
\caption{An example of the ``scan-and-update'' procedure to cancel ambient electric field using 89S state. The solid arrows indicate that the fitted center of the parabola is updated to the corresponding electrodes. The dashed arrows indicate no update is needed, as the center agrees with the previous iteration with the error bar.  
The procedure converges to a stray electric field of 
237(3) mV/cm
after multiple iterations (four iterations not shown).}
\label{fig:Iteration_Sequence_3}
\end{figure*}

\subsection{Elimination of stray electric fields}
\label{sec:Elimination of stray electric fields}



The shift in the frequency of the transition ($\delta$) as a function of applied (subscript A) and stray (subscript S) electric fields is as follows:
\begin{equation}
    \delta= -\cfrac{1}{2}\alpha(\textbf{E}_A+\textbf{E}_S)^2.
    \label{eq1}
\end{equation}
 In order to scan $\textbf{E}_A$, we apply voltages $V_i$, $i=x,y,z$, to the electrodes (Table~\ref{tab:how_to_apply_E_field}). 
 For instance, we scan $E_{A,x}$ by applying a 
 voltage $V_x$ to electrode 8 and an equal but opposite voltage $-V_x$ to electrode 4, resulting in a voltage difference of 2$V_x$ between these two electrodes. At a given $\textbf{E}_A$, we perform spectroscopy to determine the frequency shift $\delta$. We do this by scanning the frequency of the control laser and fitting the data to a Gaussian fit function (Fig.~\ref{fig:Rydberg_spectroscopy}(a-d)). We fit the $\delta$ data as a function of the applied voltage $V_i$ using the following relation (Fig.~\ref{fig:Rydberg_spectroscopy}(e))  
 \begin{equation}
    \delta= - a_i(V_{i}-V_{0,i})^2.
    \label{eq2}
\end{equation}
 We extract the center $V_{0,i}$ from these fits. We update the default voltages applied to the electrodes if the difference between the extracted value $V_{0,i}$ and the default value is statistically significant. This ``scan-and-update'' procedure (Fig.~\ref{fig:Iteration_Sequence_3}) is repeated for each axis: scan $x$, update $x$; scan $y$, update $y$; scan $z$, update $z$. Once the ``scan-and-update'' procedure is completed for all three axes, an iteration is complete. We perform such iterations until no further updates are required. Typically, the voltages converge within six such iterations.

We observe that changing the voltage on the electrodes in a manner that maintains a near-zero average voltage between consecutive points removes systematic effects arising from redistribution of charge on the glass cell walls~\cite{saskin2021building,levine2018high} (dashed arrows in Fig.~\ref{fig:Rydberg_spectroscopy}(e)). Previous studies reported that it takes up to 30~min for the charge distribution to settle~\cite{wilson2022trapping}, and UV light helps speed up this process.

To convert the applied voltages to electric fields, we average the leading coefficient $a_i$ from the fits (Eq.~\ref{eq2}) of resonant frequency $\delta$ vs $V_i$ data along each direction $i=x,y,z$. This gives us $\{a_x,a_y,a_z\}=\{51(3), 91(3), 22(1)\}$~$\text{MHz }V^{-2}$.  
The ratio of the electric field to the applied voltage $\left|E_{A,i}/V_i\right|$ is $\sqrt{2a_i/\alpha}$, where $\alpha= h \times2.94 \text{ GHz cm}^2 V^{-2}$ is the polarizability of the 89S state~\cite{vsibalic2017arc}. 

\subsection{Long term drift}
\label{sec:long term drift}

Upon determining the voltages to be applied to the electrodes to nullify the stray electric fields, we investigate the duration for which these voltages continue to nullify the ambient stray field. To track and quantify these drifts, we perform Rydberg spectroscopy (Fig.~\ref{fig:Rydberg_spectroscopy}) at specific time intervals without updating the default voltages on the electrodes. We use $ E_{\text{drift}}=\sqrt{2\sum_{i=x,y,z}a_i(V^{\textrm{new}}_{0,i}-V^{\textrm{null}}_{0,i})^2/\alpha}$ as a measure of ambient electric field drifts, where $V^{\textrm{new}}_{0,i}$ and $V^{\textrm{null}}_{0,i}$ represent the newly measured $V_{0,i}$ and the previously established $V_{0,i}$ that nullifies the ambient field, respectively. $V^{\textrm{new}}_{0,i}$ falls well within the error bars of $V^{\textrm{null}}_{0,i}$ when probed immediately after the reference scan from which $V^{\textrm{null}}_{0,i}$ was derived. Over a short-term duration of a few hours, the electric field exhibits a small drift, typically not exceeding 20~mV/cm. Day-to-day variations present a more noticeable drift, generally capped at 50~mV/cm.


\begin{figure}[ht]
\centering
\includegraphics[width=0.6\columnwidth]{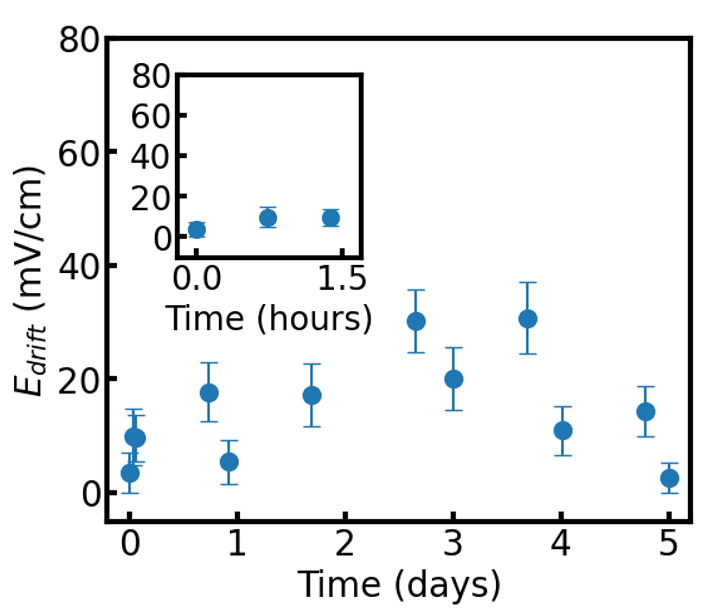}
\caption{Electric field stability measurement. The inset zooms into the first 1.5~hrs. 
}
\label{fig:E_field_drift_3}
\end{figure}

\subsection{Electric field inhomogeneity}

In our spectroscopy scans used for cancelling stray fields, we observe resonant features with Gaussian Root-Mean-Square (RMS) widths of 5~MHz. We attribute these widths to the linewidth of the cooling laser and inhomogeneous broadening caused by the magnetic field distribution in a cMOT. 
To estimate the inhomogeneous broadening caused by the applied electric fields,  we model the inhomogeneity of the DC Stark shift as~\footnote{At relatively large electric field, the shift significantly deviates from $\frac{1}{2}\alpha E^2$ (Appendix B). We fit custom polynomial expressions to capture shifts within a specific small range of electric fields.}
\begin{align}
\Delta U = \frac{1}{2}\alpha \sqrt{\overline{E^4}-\overline{E^2}^2}
\label{eq:definition of broadening}
\end{align}
with 
\begin{equation}
\overline{E^n} = \int \,dV\,e^{-\frac{x^2+y^2+z^2}{2\sigma^2}} |\mathbf{E}_A+\mathbf{E}_S|^n  /V
\label{eq:model of E field}
\end{equation}
The equation shows that the broadening effect increases with electric field strength, even for a fixed gradient.

The stray electric field ($\mathbf{E}_S$) is assumed to be constant across the atomic cloud, while the applied electric field ($\mathbf{E}_A$) is simulated using COMSOL based on the voltages applied to the corresponding electrodes.
 The volume of interest is a cube with a side length of $100~\mu$m, comparable in size to our atomic cloud, which has an RMS size of $70~\mu$m along the gradient coil axis.
 At the null total electric field point $\textbf{E}_A = -\textbf{E}_S$, our numerical calculations reveal a broadening of $\sim20$~kHz due to the applied electric field, which is robust against imperfect electrodes and atomic cloud positioning. A broadening on the order of tens of kilohertz is smaller than the Doppler width at the temperature of our atomic cloud and many other Rydberg experiments. Our simulations exhibit qualitative agreement with experimental results regarding the leading coefficients of the parabolas and additional broadening at large electric fields. Both of them are sensitive to electrodes and atomic cloud positioning errors, along with perhaps the effect from rubidium atom coating the glass cell. For a detailed discussion on broadening estimation and comparisons with experimental results, please refer to Appendix C.

\section{\label{sec:Conclusion}Conclusion}
In conclusion, we successfully designed, implemented, and characterized novel in-vacuo electrodes that can effectively cancel stray electric fields to within 10 mV/cm. This design is readily compatible with virtually any ultra-high vacuum (UHV) glass cell, making it an essential component for Rydberg experiments involving high principal quantum numbers $n$ or high angular momentum $l$. 
For example, the 89S state mixes with high $l$ states at electric fields as low as 100~mV/cm (see Appendix B).  Besides shifting the Rydberg states, electric fields may also drastically alter their pair interaction potential~\cite{de2018quantum}, even with changes as small as 20~mV/cm. Our and similar UHV glass cell systems typically have stray electric fields $\sim200$~mV/cm. Therefore, proper control of electric field environment is vital.


Our electrode design aims to achieve full electric field control through the use of eight independently controlled electrodes. Unfortunately, during the baking process, the wires of two electrodes became shorted. In the future, this problem can be avoided by routing the wires through structures that provide well-defined paths. 
Additionally, using thicker, more rigid tungsten rods may improve the agreement between the design and experimental results, which would be especially critical if the glass cell were connected to the stainless steel vacuum chamber horizontally rather than vertically, as sagging due to gravity will be more pronounced in the horizontal configuration.

\begin{acknowledgments}
We thank Steven L. Rolston, James V. Porto, Tout T. Wang, Jeff D.Thompson, and Dolev Bluvstein for sharing their in-vacuo electrode design. Our design is based on the one from Steven L. Rolston and James V. Porto. This work was supported by Purdue startup fund and AFOSR Grant FA9550-22-1-0327.
\end{acknowledgments}

\section*{Data Availability Statement}

All spectroscopy data and the CAD file of our electrode design are openly available in Zenodo at \url{https://doi.org/10.5281/zenodo.13308692}.
Additional data that support the findings of this study are available from the corresponding author upon reasonable request.

\section*{Appendix A: Control laser lock}
\begin{figure}[ht]
\centering
\includegraphics[width=0.8\columnwidth]{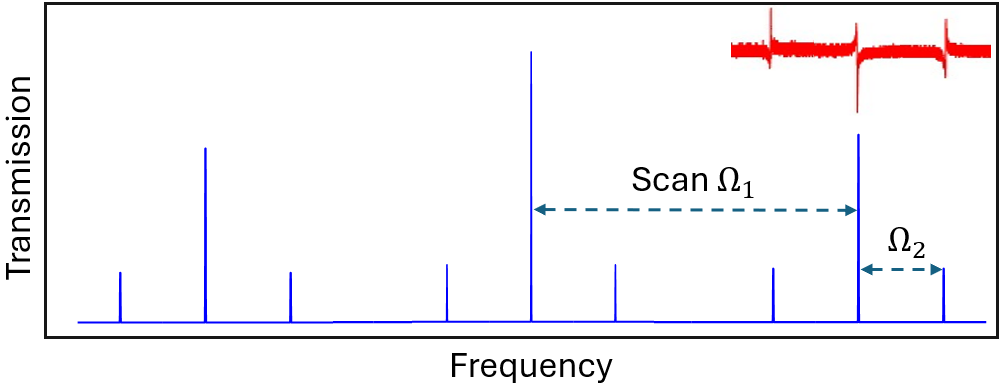}
\caption{Schematic of sidebands generated by fiber and free space EOMs. The modulation frequencies are $\Omega_1$ for the fiber EOM and $\Omega_2=25$~MHz for the free-space EOM. Locking to the -1 (+1) order sideband of the fiber EOM allows $\Omega_1$ to control the carrier frequency, with increases (decreases) in $\Omega_1$ shifting the frequency of the control beam upwards. The red curve is a sketch of the PDH error signal.}
\label{fig:control_lock_error_transmission}
\end{figure}

The control beam is derived from a frequency-doubled titanium sapphire (Ti:Saph) laser (SOLSTIS ECD-X, M Squared) and stabilized to an ultra-low-expansion (ULE) cavity (6010-4, Stable Laser Systems) using a Pound–Drever–Hall (PDH) lock. The 958~nm fundamental light passes through a fiber electro-optic modulator (EOM) (PM-0S5-10-PFA-PFA-960, EOSpace) and then a free-space EOM (PM7-NIR$\_$25, QUBIG). The fiber EOM has a large bandwidth up to $10$~GHz. Since the resonance frequency of the cavity is not tunable within its free spectral range (1.5~GHz), either the +1 or -1 order sideband ($20-730$~MHz) of the fiber EOM is locked to the cavity (Fig.~\ref{fig:control_lock_error_transmission}). For spectroscopy, we sweep the frequency of the radio-frequency sinusoidal signal driving the fiber EOM. We have multiplied the factor of 2 to account for the frequency doubling, as well as the sign difference when locking to either the +1 or -1 order sidebands, in Figs.~\ref{fig:Rydberg_spectroscopy}, \ref{fig:Iteration_Sequence_3} and~\ref{fig:86S}. The free-space EOM is resonant at 25~MHz, which is the modulation frequency of the PDH lock.

\section*{Appendix B: Stark map }

\begin{figure}[ht]
\centering
\includegraphics[width=0.8\columnwidth]{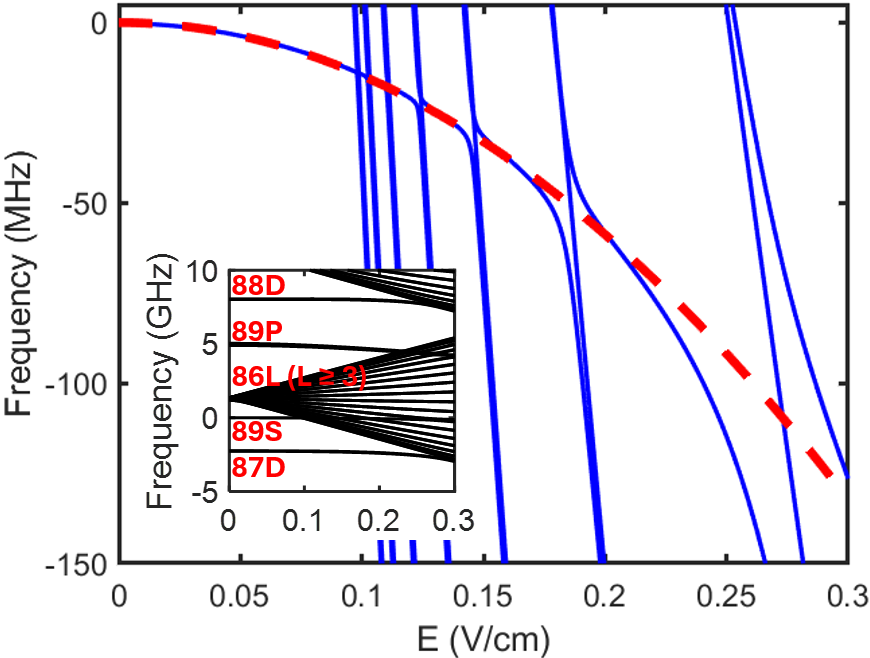}
\caption{Stark map near 89S state. The dashed red line is $\frac{1}{2}\alpha E^2$, with $\alpha= 2.94 \text{ GHz cm}^2 V^{-2}$. The inset zooms out to a larger range of frequency.}
\label{fig:StarkMap}
\end{figure}

\section*{Appendix C: Electric field inhomogeneity}
\begin{figure*}
\centering
\includegraphics[width=0.65\textwidth]{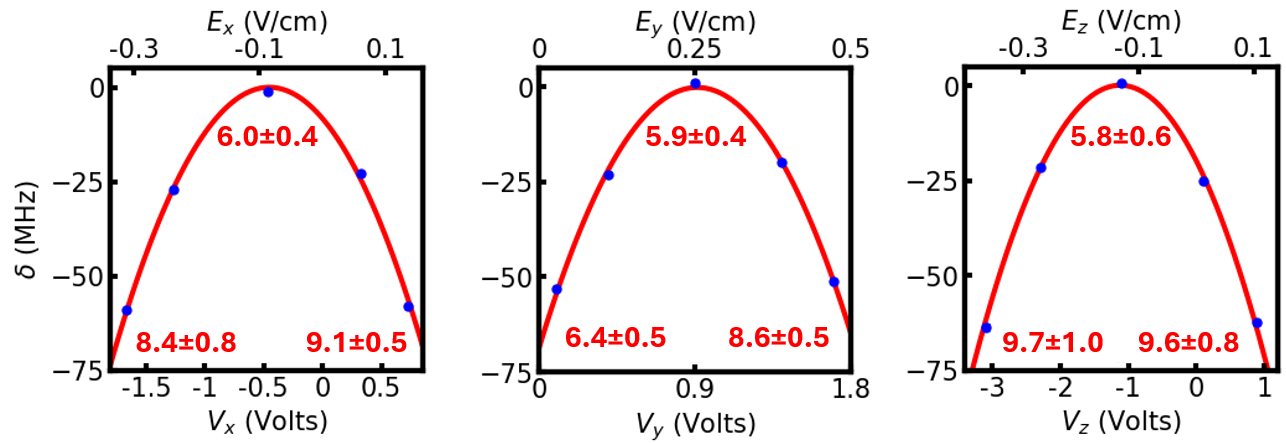}
\caption{Resonant frequency vs electric fields for the 86S state. The numbers in red indicate the observed Gaussian RMS  widths, in MHz, of selected spectroscopy dips. We apply larger electric field to broaden the resonant features, whereas for cancelling the stray fields, we confine ourselves to the range within which the shifts agree well with $\frac{1}{2}\alpha E^2$.}
\label{fig:86S}
\end{figure*}
For the 86S state, COMSOL simulation predicts that the leading coefficients of the parabolas are $\{a_x,a_y,a_z\}=\{40, 188, 27\}$~$\text{MHz }V^{-2}$, whereas the measured results are $\{a_x,a_y,a_z\}=\{40(1), 77(2), 16(0)\}$~$\text{MHz }V^{-2}$.
To estimate inhomogeneous broadening caused by the applied electric fields, we apply the same voltages to the electrodes in the simulation as in the experiment. We assume after we cancel the stray electric field, the total field is zero at the geometric center. In other words, the stray field is assumed to have the same magnitude and opposite sign as the simulated electric field resulting from the voltages apply at the apex of the three parabolas. For example, in Fig.~\ref{fig:86S}, the stray field is inferred to be $\mathbf{E}_S= 88\mathbf{e}_x-363\mathbf{e}_y-169\mathbf{e}_z$ mV/cm across the entire volume of interest, a cube with a side length of 0.1~mm, comparable to the size of our atomic cloud. 
At $\mathbf{E}_A=-311\mathbf{e}_x+364\mathbf{e}_y+169\mathbf{e}_z$ mV/cm, we observe the width of the resonant feature is 8.4~MHz (first panel in Fig.~\ref{fig:86S}), broadened from that of the zero total electric field point, which is 6~MHz. This is consistent with the results from our simulation, which predicts the broadening to be enlarged to 7.6~MHz. 
All the six labeled widths of the resonant features in Fig.~\ref{fig:86S} have similarly qualitative agreements with the results from our simulation. By displacing electrodes and/or the atomic cloud, we can improve agreements of certain data points. However, we did not pursue an exact match for all electric field-to-voltage ratios and resonant dip widths at large electric fields, due to the process's time-consuming nature and limited informative value.


\vspace{7mm}
\section*{References}
\vspace{-3.7mm}
\bibliography{bibliography}

\begin{thebibliography}{25}%
\makeatletter
\providecommand \@ifxundefined [1]{%
 \@ifx{#1\undefined}
}%
\providecommand \@ifnum [1]{%
 \ifnum #1\expandafter \@firstoftwo
 \else \expandafter \@secondoftwo
 \fi
}%
\providecommand \@ifx [1]{%
 \ifx #1\expandafter \@firstoftwo
 \else \expandafter \@secondoftwo
 \fi
}%
\providecommand \natexlab [1]{#1}%
\providecommand \enquote  [1]{``#1''}%
\providecommand \bibnamefont  [1]{#1}%
\providecommand \bibfnamefont [1]{#1}%
\providecommand \citenamefont [1]{#1}%
\providecommand \href@noop [0]{\@secondoftwo}%
\providecommand \href [0]{\begingroup \@sanitize@url \@href}%
\providecommand \@href[1]{\@@startlink{#1}\@@href}%
\providecommand \@@href[1]{\endgroup#1\@@endlink}%
\providecommand \@sanitize@url [0]{\catcode `\\12\catcode `\$12\catcode `\&12\catcode `\#12\catcode `\^12\catcode `\_12\catcode `\%12\relax}%
\providecommand \@@startlink[1]{}%
\providecommand \@@endlink[0]{}%
\providecommand \url  [0]{\begingroup\@sanitize@url \@url }%
\providecommand \@url [1]{\endgroup\@href {#1}{\urlprefix }}%
\providecommand \urlprefix  [0]{URL }%
\providecommand \Eprint [0]{\href }%
\providecommand \doibase [0]{http://dx.doi.org/}%
\providecommand \selectlanguage [0]{\@gobble}%
\providecommand \bibinfo  [0]{\@secondoftwo}%
\providecommand \bibfield  [0]{\@secondoftwo}%
\providecommand \translation [1]{[#1]}%
\providecommand \BibitemOpen [0]{}%
\providecommand \bibitemStop [0]{}%
\providecommand \bibitemNoStop [0]{.\EOS\space}%
\providecommand \EOS [0]{\spacefactor3000\relax}%
\providecommand \BibitemShut  [1]{\csname bibitem#1\endcsname}%
\let\auto@bib@innerbib\@empty
\bibitem [{\citenamefont {Morgado}\ and\ \citenamefont {Whitlock}(2021)}]{Morgado2021}%
  \BibitemOpen
  \bibfield  {author} {\bibinfo {author} {\bibfnamefont {M.}~\bibnamefont {Morgado}}\ and\ \bibinfo {author} {\bibfnamefont {S.}~\bibnamefont {Whitlock}},\ }\bibfield  {title} {\enquote {\bibinfo {title} {{Quantum simulation and computing with Rydberg-interacting qubits}},}\ }\href {\doibase 10.1116/5.0036562} {\bibfield  {journal} {\bibinfo  {journal} {AVS Quantum Science}\ }\textbf {\bibinfo {volume} {3}} (\bibinfo {year} {2021}),\ 10.1116/5.0036562},\ \Eprint {http://arxiv.org/abs/2011.03031} {arXiv:2011.03031} \BibitemShut {NoStop}%
\bibitem [{\citenamefont {Browaeys}\ and\ \citenamefont {Lahaye}(2020)}]{browaeys2020many}%
  \BibitemOpen
  \bibfield  {author} {\bibinfo {author} {\bibfnamefont {A.}~\bibnamefont {Browaeys}}\ and\ \bibinfo {author} {\bibfnamefont {T.}~\bibnamefont {Lahaye}},\ }\bibfield  {title} {\enquote {\bibinfo {title} {Many-body physics with individually controlled rydberg atoms},}\ }\href@noop {} {\bibfield  {journal} {\bibinfo  {journal} {Nature Physics}\ }\textbf {\bibinfo {volume} {16}},\ \bibinfo {pages} {132--142} (\bibinfo {year} {2020})}\BibitemShut {NoStop}%
\bibitem [{\citenamefont {Bluvstein}\ \emph {et~al.}(2024)\citenamefont {Bluvstein}, \citenamefont {Evered}, \citenamefont {Geim}, \citenamefont {Li}, \citenamefont {Zhou}, \citenamefont {Manovitz}, \citenamefont {Ebadi}, \citenamefont {Cain}, \citenamefont {Kalinowski}, \citenamefont {Hangleiter} \emph {et~al.}}]{bluvstein2024logical}%
  \BibitemOpen
  \bibfield  {author} {\bibinfo {author} {\bibfnamefont {D.}~\bibnamefont {Bluvstein}}, \bibinfo {author} {\bibfnamefont {S.~J.}\ \bibnamefont {Evered}}, \bibinfo {author} {\bibfnamefont {A.~A.}\ \bibnamefont {Geim}}, \bibinfo {author} {\bibfnamefont {S.~H.}\ \bibnamefont {Li}}, \bibinfo {author} {\bibfnamefont {H.}~\bibnamefont {Zhou}}, \bibinfo {author} {\bibfnamefont {T.}~\bibnamefont {Manovitz}}, \bibinfo {author} {\bibfnamefont {S.}~\bibnamefont {Ebadi}}, \bibinfo {author} {\bibfnamefont {M.}~\bibnamefont {Cain}}, \bibinfo {author} {\bibfnamefont {M.}~\bibnamefont {Kalinowski}}, \bibinfo {author} {\bibfnamefont {D.}~\bibnamefont {Hangleiter}},  \emph {et~al.},\ }\bibfield  {title} {\enquote {\bibinfo {title} {Logical quantum processor based on reconfigurable atom arrays},}\ }\href@noop {} {\bibfield  {journal} {\bibinfo  {journal} {Nature}\ }\textbf {\bibinfo {volume} {626}},\ \bibinfo {pages} {58--65} (\bibinfo {year} {2024})}\BibitemShut {NoStop}%
\bibitem [{\citenamefont {Firstenberg}, \citenamefont {Adams},\ and\ \citenamefont {Hofferberth}(2016)}]{firstenberg2016nonlinear}%
  \BibitemOpen
  \bibfield  {author} {\bibinfo {author} {\bibfnamefont {O.}~\bibnamefont {Firstenberg}}, \bibinfo {author} {\bibfnamefont {C.~S.}\ \bibnamefont {Adams}}, \ and\ \bibinfo {author} {\bibfnamefont {S.}~\bibnamefont {Hofferberth}},\ }\bibfield  {title} {\enquote {\bibinfo {title} {Nonlinear quantum optics mediated by rydberg interactions},}\ }\href@noop {} {\bibfield  {journal} {\bibinfo  {journal} {Journal of Physics B: Atomic, Molecular and Optical Physics}\ }\textbf {\bibinfo {volume} {49}},\ \bibinfo {pages} {152003} (\bibinfo {year} {2016})}\BibitemShut {NoStop}%
\bibitem [{\citenamefont {Shao}\ \emph {et~al.}(2024)\citenamefont {Shao}, \citenamefont {Su}, \citenamefont {Li}, \citenamefont {Nath}, \citenamefont {Wu},\ and\ \citenamefont {Li}}]{shao2024rydberg}%
  \BibitemOpen
  \bibfield  {author} {\bibinfo {author} {\bibfnamefont {X.-Q.}\ \bibnamefont {Shao}}, \bibinfo {author} {\bibfnamefont {S.-L.}\ \bibnamefont {Su}}, \bibinfo {author} {\bibfnamefont {L.}~\bibnamefont {Li}}, \bibinfo {author} {\bibfnamefont {R.}~\bibnamefont {Nath}}, \bibinfo {author} {\bibfnamefont {J.-H.}\ \bibnamefont {Wu}}, \ and\ \bibinfo {author} {\bibfnamefont {W.}~\bibnamefont {Li}},\ }\bibfield  {title} {\enquote {\bibinfo {title} {Rydberg superatoms: An artificial quantum system for quantum information processing and quantum optics},}\ }\href@noop {} {\bibfield  {journal} {\bibinfo  {journal} {Applied Physics Reviews}\ }\textbf {\bibinfo {volume} {11}} (\bibinfo {year} {2024})}\BibitemShut {NoStop}%
\bibitem [{\citenamefont {Meyer}\ \emph {et~al.}(2020)\citenamefont {Meyer}, \citenamefont {Castillo}, \citenamefont {Cox},\ and\ \citenamefont {Kunz}}]{meyer2020assessment}%
  \BibitemOpen
  \bibfield  {author} {\bibinfo {author} {\bibfnamefont {D.~H.}\ \bibnamefont {Meyer}}, \bibinfo {author} {\bibfnamefont {Z.~A.}\ \bibnamefont {Castillo}}, \bibinfo {author} {\bibfnamefont {K.~C.}\ \bibnamefont {Cox}}, \ and\ \bibinfo {author} {\bibfnamefont {P.~D.}\ \bibnamefont {Kunz}},\ }\bibfield  {title} {\enquote {\bibinfo {title} {Assessment of rydberg atoms for wideband electric field sensing},}\ }\href@noop {} {\bibfield  {journal} {\bibinfo  {journal} {Journal of Physics B: Atomic, Molecular and Optical Physics}\ }\textbf {\bibinfo {volume} {53}},\ \bibinfo {pages} {034001} (\bibinfo {year} {2020})}\BibitemShut {NoStop}%
\bibitem [{\citenamefont {Fancher}\ \emph {et~al.}(2021)\citenamefont {Fancher}, \citenamefont {Scherer}, \citenamefont {John},\ and\ \citenamefont {Marlow}}]{fancher2021rydberg}%
  \BibitemOpen
  \bibfield  {author} {\bibinfo {author} {\bibfnamefont {C.~T.}\ \bibnamefont {Fancher}}, \bibinfo {author} {\bibfnamefont {D.~R.}\ \bibnamefont {Scherer}}, \bibinfo {author} {\bibfnamefont {M.~C.~S.}\ \bibnamefont {John}}, \ and\ \bibinfo {author} {\bibfnamefont {B.~L.~S.}\ \bibnamefont {Marlow}},\ }\bibfield  {title} {\enquote {\bibinfo {title} {Rydberg atom electric field sensors for communications and sensing},}\ }\href@noop {} {\bibfield  {journal} {\bibinfo  {journal} {IEEE Transactions on Quantum Engineering}\ }\textbf {\bibinfo {volume} {2}},\ \bibinfo {pages} {1--13} (\bibinfo {year} {2021})}\BibitemShut {NoStop}%
\bibitem [{\citenamefont {Saffman}, \citenamefont {Walker},\ and\ \citenamefont {M\o{}lmer}(2010)}]{RevModPhys.82.2313}%
  \BibitemOpen
  \bibfield  {author} {\bibinfo {author} {\bibfnamefont {M.}~\bibnamefont {Saffman}}, \bibinfo {author} {\bibfnamefont {T.~G.}\ \bibnamefont {Walker}}, \ and\ \bibinfo {author} {\bibfnamefont {K.}~\bibnamefont {M\o{}lmer}},\ }\bibfield  {title} {\enquote {\bibinfo {title} {Quantum information with rydberg atoms},}\ }\href {\doibase 10.1103/RevModPhys.82.2313} {\bibfield  {journal} {\bibinfo  {journal} {Rev. Mod. Phys.}\ }\textbf {\bibinfo {volume} {82}},\ \bibinfo {pages} {2313--2363} (\bibinfo {year} {2010})}\BibitemShut {NoStop}%
\bibitem [{\citenamefont {Evered}\ \emph {et~al.}(2023)\citenamefont {Evered}, \citenamefont {Bluvstein}, \citenamefont {Kalinowski}, \citenamefont {Ebadi}, \citenamefont {Manovitz}, \citenamefont {Zhou}, \citenamefont {Li}, \citenamefont {Geim}, \citenamefont {Wang}, \citenamefont {Maskara} \emph {et~al.}}]{evered2023high}%
  \BibitemOpen
  \bibfield  {author} {\bibinfo {author} {\bibfnamefont {S.~J.}\ \bibnamefont {Evered}}, \bibinfo {author} {\bibfnamefont {D.}~\bibnamefont {Bluvstein}}, \bibinfo {author} {\bibfnamefont {M.}~\bibnamefont {Kalinowski}}, \bibinfo {author} {\bibfnamefont {S.}~\bibnamefont {Ebadi}}, \bibinfo {author} {\bibfnamefont {T.}~\bibnamefont {Manovitz}}, \bibinfo {author} {\bibfnamefont {H.}~\bibnamefont {Zhou}}, \bibinfo {author} {\bibfnamefont {S.~H.}\ \bibnamefont {Li}}, \bibinfo {author} {\bibfnamefont {A.~A.}\ \bibnamefont {Geim}}, \bibinfo {author} {\bibfnamefont {T.~T.}\ \bibnamefont {Wang}}, \bibinfo {author} {\bibfnamefont {N.}~\bibnamefont {Maskara}},  \emph {et~al.},\ }\bibfield  {title} {\enquote {\bibinfo {title} {High-fidelity parallel entangling gates on a neutral-atom quantum computer},}\ }\href@noop {} {\bibfield  {journal} {\bibinfo  {journal} {Nature}\ }\textbf {\bibinfo {volume} {622}},\ \bibinfo {pages} {268--272} (\bibinfo {year} {2023})}\BibitemShut {NoStop}%
\bibitem [{\citenamefont {Peyronel}(2013)}]{peyronel2013quantum}%
  \BibitemOpen
  \bibfield  {author} {\bibinfo {author} {\bibfnamefont {T.~M.~M.}\ \bibnamefont {Peyronel}},\ }\emph {\bibinfo {title} {Quantum nonlinear optics using cold atomic ensembles}},\ \href@noop {} {Ph.D. thesis},\ \bibinfo  {school} {Massachusetts Institute of Technology} (\bibinfo {year} {2013})\BibitemShut {NoStop}%
\bibitem [{\citenamefont {de~L{\'e}s{\'e}leuc}(2018)}]{de2018quantum}%
  \BibitemOpen
  \bibfield  {author} {\bibinfo {author} {\bibfnamefont {S.}~\bibnamefont {de~L{\'e}s{\'e}leuc}},\ }\emph {\bibinfo {title} {Quantum simulation of spin models with assembled arrays of Rydberg atoms}},\ \href@noop {} {Ph.D. thesis},\ \bibinfo  {school} {Ecole Polytechnique} (\bibinfo {year} {2018})\BibitemShut {NoStop}%
\bibitem [{\citenamefont {Ornelas~Huerta}(2020)}]{ornelas2020experiments}%
  \BibitemOpen
  \bibfield  {author} {\bibinfo {author} {\bibfnamefont {D.~P.}\ \bibnamefont {Ornelas~Huerta}},\ }\emph {\bibinfo {title} {Experiments with strongly-interacting Rydberg atoms}},\ \href@noop {} {Ph.D. thesis},\ \bibinfo  {school} {Joint Quantum Institute, National Institute of Standards and Technology and University of Maryland College Park} (\bibinfo {year} {2020})\BibitemShut {NoStop}%
\bibitem [{\citenamefont {Mirgorodskiy}(2017)}]{mirgorodskiy2017storage}%
  \BibitemOpen
  \bibfield  {author} {\bibinfo {author} {\bibfnamefont {I.}~\bibnamefont {Mirgorodskiy}},\ }\emph {\bibinfo {title} {Storage and propagation of Rydberg polaritons in a cold atomic medium}},\ \href@noop {} {Ph.D. thesis},\ \bibinfo  {school} {Universität Stuttgart} (\bibinfo {year} {2017})\BibitemShut {NoStop}%
\bibitem [{\citenamefont {Lorenz}(2021)}]{lorenz2021rydberg}%
  \BibitemOpen
  \bibfield  {author} {\bibinfo {author} {\bibfnamefont {N.}~\bibnamefont {Lorenz}},\ }\emph {\bibinfo {title} {A Rydberg tweezer platform with potassium atoms}},\ \href@noop {} {Ph.D. thesis},\ \bibinfo  {school} {Ludwig-Maximilians-Universität} (\bibinfo {year} {2021})\BibitemShut {NoStop}%
\bibitem [{\citenamefont {Anand}\ \emph {et~al.}(2024)\citenamefont {Anand}, \citenamefont {Bradley}, \citenamefont {White}, \citenamefont {Ramesh}, \citenamefont {Singh},\ and\ \citenamefont {Bernien}}]{anand2024dual}%
  \BibitemOpen
  \bibfield  {author} {\bibinfo {author} {\bibfnamefont {S.}~\bibnamefont {Anand}}, \bibinfo {author} {\bibfnamefont {C.~E.}\ \bibnamefont {Bradley}}, \bibinfo {author} {\bibfnamefont {R.}~\bibnamefont {White}}, \bibinfo {author} {\bibfnamefont {V.}~\bibnamefont {Ramesh}}, \bibinfo {author} {\bibfnamefont {K.}~\bibnamefont {Singh}}, \ and\ \bibinfo {author} {\bibfnamefont {H.}~\bibnamefont {Bernien}},\ }\href {https://arxiv.org/abs/2401.10325} {\enquote {\bibinfo {title} {A dual-species rydberg array},}\ } (\bibinfo {year} {2024}),\ \Eprint {http://arxiv.org/abs/2401.10325} {arXiv:2401.10325 [quant-ph]} \BibitemShut {NoStop}%
\bibitem [{\citenamefont {Saffman}, \citenamefont {Noel},\ and\ \citenamefont {Hughes}(2024)}]{saffmanelelectricfieldcontrolpatent}%
  \BibitemOpen
  \bibfield  {author} {\bibinfo {author} {\bibfnamefont {M.}~\bibnamefont {Saffman}}, \bibinfo {author} {\bibfnamefont {T.~W.}\ \bibnamefont {Noel}}, \ and\ \bibinfo {author} {\bibfnamefont {S.~M.}\ \bibnamefont {Hughes}},\ }\href {https://patents.google.com/patent/US11997780B2/en} {\enquote {\bibinfo {title} {Vacuum cell with electric-field control},}\ } (\bibinfo {year} {2024}),\ \bibinfo {note} {{U.S. Patent No.} 11997780B2}\BibitemShut {NoStop}%
\bibitem [{\citenamefont {Wilson}\ \emph {et~al.}(2022)\citenamefont {Wilson}, \citenamefont {Saskin}, \citenamefont {Meng}, \citenamefont {Ma}, \citenamefont {Dilip}, \citenamefont {Burgers},\ and\ \citenamefont {Thompson}}]{wilson2022trapping}%
  \BibitemOpen
  \bibfield  {author} {\bibinfo {author} {\bibfnamefont {J.}~\bibnamefont {Wilson}}, \bibinfo {author} {\bibfnamefont {S.}~\bibnamefont {Saskin}}, \bibinfo {author} {\bibfnamefont {Y.}~\bibnamefont {Meng}}, \bibinfo {author} {\bibfnamefont {S.}~\bibnamefont {Ma}}, \bibinfo {author} {\bibfnamefont {R.}~\bibnamefont {Dilip}}, \bibinfo {author} {\bibfnamefont {A.}~\bibnamefont {Burgers}}, \ and\ \bibinfo {author} {\bibfnamefont {J.}~\bibnamefont {Thompson}},\ }\bibfield  {title} {\enquote {\bibinfo {title} {Trapping alkaline earth rydberg atoms optical tweezer arrays},}\ }\href@noop {} {\bibfield  {journal} {\bibinfo  {journal} {Physical Review Letters}\ }\textbf {\bibinfo {volume} {128}},\ \bibinfo {pages} {033201} (\bibinfo {year} {2022})}\BibitemShut {NoStop}%
\bibitem [{\citenamefont {Wilson}(2022)}]{wilson2022new}%
  \BibitemOpen
  \bibfield  {author} {\bibinfo {author} {\bibfnamefont {J.}~\bibnamefont {Wilson}},\ }\emph {\bibinfo {title} {New tools for quantum science in Yb Rydberg atom arrays}},\ \href@noop {} {Ph.D. thesis},\ \bibinfo  {school} {Princeton University} (\bibinfo {year} {2022})\BibitemShut {NoStop}%
\bibitem [{\citenamefont {Archimi}\ \emph {et~al.}(2022)\citenamefont {Archimi}, \citenamefont {Ceccanti}, \citenamefont {Distefano}, \citenamefont {Di~Virgilio}, \citenamefont {Franco}, \citenamefont {Greco}, \citenamefont {Simonelli}, \citenamefont {Arimondo}, \citenamefont {Ciampini},\ and\ \citenamefont {Morsch}}]{archimi2022measurements}%
  \BibitemOpen
  \bibfield  {author} {\bibinfo {author} {\bibfnamefont {M.}~\bibnamefont {Archimi}}, \bibinfo {author} {\bibfnamefont {M.}~\bibnamefont {Ceccanti}}, \bibinfo {author} {\bibfnamefont {M.}~\bibnamefont {Distefano}}, \bibinfo {author} {\bibfnamefont {L.}~\bibnamefont {Di~Virgilio}}, \bibinfo {author} {\bibfnamefont {R.}~\bibnamefont {Franco}}, \bibinfo {author} {\bibfnamefont {A.}~\bibnamefont {Greco}}, \bibinfo {author} {\bibfnamefont {C.}~\bibnamefont {Simonelli}}, \bibinfo {author} {\bibfnamefont {E.}~\bibnamefont {Arimondo}}, \bibinfo {author} {\bibfnamefont {D.}~\bibnamefont {Ciampini}}, \ and\ \bibinfo {author} {\bibfnamefont {O.}~\bibnamefont {Morsch}},\ }\bibfield  {title} {\enquote {\bibinfo {title} {Measurements of blackbody-radiation-induced transition rates between high-lying s, p, and d rydberg levels},}\ }\href@noop {} {\bibfield  {journal} {\bibinfo  {journal} {Physical Review A}\ }\textbf {\bibinfo {volume} {105}},\ \bibinfo {pages} {063104} (\bibinfo {year} {2022})}\BibitemShut {NoStop}%
\bibitem [{\citenamefont {Levine}\ \emph {et~al.}(2018)\citenamefont {Levine}, \citenamefont {Keesling}, \citenamefont {Omran}, \citenamefont {Bernien}, \citenamefont {Schwartz}, \citenamefont {Zibrov}, \citenamefont {Endres}, \citenamefont {Greiner}, \citenamefont {Vuleti{\'c}},\ and\ \citenamefont {Lukin}}]{levine2018high}%
  \BibitemOpen
  \bibfield  {author} {\bibinfo {author} {\bibfnamefont {H.}~\bibnamefont {Levine}}, \bibinfo {author} {\bibfnamefont {A.}~\bibnamefont {Keesling}}, \bibinfo {author} {\bibfnamefont {A.}~\bibnamefont {Omran}}, \bibinfo {author} {\bibfnamefont {H.}~\bibnamefont {Bernien}}, \bibinfo {author} {\bibfnamefont {S.}~\bibnamefont {Schwartz}}, \bibinfo {author} {\bibfnamefont {A.~S.}\ \bibnamefont {Zibrov}}, \bibinfo {author} {\bibfnamefont {M.}~\bibnamefont {Endres}}, \bibinfo {author} {\bibfnamefont {M.}~\bibnamefont {Greiner}}, \bibinfo {author} {\bibfnamefont {V.}~\bibnamefont {Vuleti{\'c}}}, \ and\ \bibinfo {author} {\bibfnamefont {M.~D.}\ \bibnamefont {Lukin}},\ }\bibfield  {title} {\enquote {\bibinfo {title} {High-fidelity control and entanglement of rydberg-atom qubits},}\ }\href@noop {} {\bibfield  {journal} {\bibinfo  {journal} {Physical Review Letters}\ }\textbf {\bibinfo {volume} {121}},\ \bibinfo {pages} {123603} (\bibinfo {year} {2018})}\BibitemShut {NoStop}%
\bibitem [{\citenamefont {Torralbo-Campo}\ \emph {et~al.}(2015)\citenamefont {Torralbo-Campo}, \citenamefont {Bruce}, \citenamefont {Smirne},\ and\ \citenamefont {Cassettari}}]{torralbo2015light}%
  \BibitemOpen
  \bibfield  {author} {\bibinfo {author} {\bibfnamefont {L.}~\bibnamefont {Torralbo-Campo}}, \bibinfo {author} {\bibfnamefont {G.~D.}\ \bibnamefont {Bruce}}, \bibinfo {author} {\bibfnamefont {G.}~\bibnamefont {Smirne}}, \ and\ \bibinfo {author} {\bibfnamefont {D.}~\bibnamefont {Cassettari}},\ }\bibfield  {title} {\enquote {\bibinfo {title} {Light-induced atomic desorption in a compact system for ultracold atoms},}\ }\href@noop {} {\bibfield  {journal} {\bibinfo  {journal} {Scientific Reports}\ }\textbf {\bibinfo {volume} {5}},\ \bibinfo {pages} {14729} (\bibinfo {year} {2015})}\BibitemShut {NoStop}%
\bibitem [{\citenamefont {Klempt}\ \emph {et~al.}(2006)\citenamefont {Klempt}, \citenamefont {Van~Zoest}, \citenamefont {Henninger}, \citenamefont {Rasel}, \citenamefont {Ertmer}, \citenamefont {Arlt} \emph {et~al.}}]{klempt2006ultraviolet}%
  \BibitemOpen
  \bibfield  {author} {\bibinfo {author} {\bibfnamefont {C.}~\bibnamefont {Klempt}}, \bibinfo {author} {\bibfnamefont {T.}~\bibnamefont {Van~Zoest}}, \bibinfo {author} {\bibfnamefont {T.}~\bibnamefont {Henninger}}, \bibinfo {author} {\bibfnamefont {E.}~\bibnamefont {Rasel}}, \bibinfo {author} {\bibfnamefont {W.}~\bibnamefont {Ertmer}}, \bibinfo {author} {\bibfnamefont {J.}~\bibnamefont {Arlt}},  \emph {et~al.},\ }\bibfield  {title} {\enquote {\bibinfo {title} {Ultraviolet light-induced atom desorption for large rubidium and potassium magneto-optical traps},}\ }\href@noop {} {\bibfield  {journal} {\bibinfo  {journal} {Physical Review A}\ }\textbf {\bibinfo {volume} {73}},\ \bibinfo {pages} {013410} (\bibinfo {year} {2006})}\BibitemShut {NoStop}%
\bibitem [{\citenamefont {Saskin}(2021)}]{saskin2021building}%
  \BibitemOpen
  \bibfield  {author} {\bibinfo {author} {\bibfnamefont {S.}~\bibnamefont {Saskin}},\ }\emph {\bibinfo {title} {Building Quantum Systems with Ytterbium Rydberg Arrays}},\ \href@noop {} {Ph.D. thesis},\ \bibinfo  {school} {Princeton University} (\bibinfo {year} {2021})\BibitemShut {NoStop}%
\bibitem [{\citenamefont {{\v{S}}ibali{\'c}}\ \emph {et~al.}(2017)\citenamefont {{\v{S}}ibali{\'c}}, \citenamefont {Pritchard}, \citenamefont {Adams},\ and\ \citenamefont {Weatherill}}]{vsibalic2017arc}%
  \BibitemOpen
  \bibfield  {author} {\bibinfo {author} {\bibfnamefont {N.}~\bibnamefont {{\v{S}}ibali{\'c}}}, \bibinfo {author} {\bibfnamefont {J.~D.}\ \bibnamefont {Pritchard}}, \bibinfo {author} {\bibfnamefont {C.~S.}\ \bibnamefont {Adams}}, \ and\ \bibinfo {author} {\bibfnamefont {K.~J.}\ \bibnamefont {Weatherill}},\ }\bibfield  {title} {\enquote {\bibinfo {title} {Arc: An open-source library for calculating properties of alkali rydberg atoms},}\ }\href@noop {} {\bibfield  {journal} {\bibinfo  {journal} {Computer Physics Communications}\ }\textbf {\bibinfo {volume} {220}},\ \bibinfo {pages} {319--331} (\bibinfo {year} {2017})}\BibitemShut {NoStop}%
\bibitem [{Note1()}]{Note1}%
  \BibitemOpen
  \bibinfo {note} {At relatively large electric field, the shift significantly deviates from $\protect \frac {1}{2}\alpha E^2$ (Appendix B). We fit custom polynomial expressions to capture shifts within a specific small range of electric fields.}\BibitemShut {Stop}%
\end{thebibliography}%

\end{document}